\documentclass[aps,pra,twocolumn,showpacs,superscriptaddress,showkeys,nofootinbib,longbibliography,floatfix]{revtex4-1}

\usepackage[utf8]{inputenc}
\usepackage[T1]{fontenc}

\usepackage{multirow}	
\usepackage{enumerate}	
\usepackage{Tabbing}	
\usepackage{relsize}	
\usepackage{textcomp}	
\usepackage{soul}	

\usepackage{amsmath}
\usepackage{amssymb}
\usepackage{bm}		

\usepackage{hyperref}
\usepackage{url}
\usepackage[capitalise]{cleveref}	

\usepackage{dcolumn}	
\usepackage{longtable}	
\usepackage{booktabs}	

\usepackage[pdftex]{graphicx}
\usepackage{epstopdf}	
\usepackage{rotating}	
\usepackage{float}	

\usepackage{listings}

\usepackage[dvipsnames]{xcolor}
\usepackage{color}
\usepackage{textcomp}

\definecolor{listinggray}{gray}{0.9}
\definecolor{lbcolor}{rgb}{0.9,0.9,0.9}

\newcommand{\ket}[1]{\ensuremath{\left|#1\right\rangle}}
\newcommand{\average}[1]{\ensuremath{\left\langle#1\right\rangle}}

\newcommand{\bracket}[2]{\ensuremath{\left\langle#1 \vphantom{#2}\right| \left. #2 \vphantom{#1}\right\rangle}}
\newcommand{\matrixel}[3]{\ensuremath{\left\langle #1 \vphantom{#2#3} \right| #2 \left| #3
\newcommand{\angstrom}{\text{\normalfont\AA}}
\vphantom{#1#2} \right\rangle}}

\newcommand{\AN}[2]{
\ensuremath{\hat{c}^{} _{{#1}
\ifnum#2=1 \uparrow
\else
\ifnum#2=-1 \downarrow
\else
\ifnum#2=2 \sigma
\else
\ifnum#2=-2 \bar{\sigma}
\else
\ifnum#2=3 \sigma '
\fi
\fi
\fi
\fi
\fi
}}
}
\newcommand{\CR}[2]{
\ensuremath{\hat{c}^\dagger _{{#1}
\ifnum#2=1 \uparrow
\else
\ifnum#2=-1 \downarrow
\else
\ifnum#2=2 \sigma
\else
\ifnum#2=-2 \bar{\sigma}
\else
\ifnum#2=3 \sigma '
\fi
\fi
\fi
\fi
\fi
}}
}
\newcommand{\NUM}[2]{
\ensuremath{\hat{n}^{} _{{#1}
\ifnum#2=1 \uparrow
\else
\ifnum#2=-1 \downarrow
\else
\ifnum#2=2 \sigma
\else
\ifnum#2=-2 \bar{\sigma}
\else
\ifnum#2=3 \sigma '
\fi
\fi
\fi
\fi
\fi
}}
}

\renewcommand{\vec}[1]{\ensuremath{\mathbf{#1}}}

\definecolor{newRed}{RGB}{200,0,0}

\definecolor{newGreen}{RGB}{0,100,0}

\begin{document}


\title{Atomization of correlated molecular-hydrogen chain: \\ A fully microscopic Variational Monte-Carlo solution}

\author{Andrzej Biborski}
\email {andrzej.biborski@agh.edu.pl}
\affiliation {Academic Centre for Materials and Nanotechnology,
AGH University of Science and Technology,
al. Mickiewicza 30, PL-30-059 Krak\'ow, Poland}

\author{Andrzej P. K\k{a}dzielawa}
\email{kadzielawa@th.if.uj.edu.pl}
\affiliation{Marian Smoluchowski Institute of Physics, Jagiellonian University, 
ulica \L{}ojasiewicza 11, PL-30-348 Krak\'ow, Poland}

\author{J\'{o}zef Spa\l{}ek}
\email{jozef.spalek@uj.edu.pl}
\affiliation{Marian Smoluchowski Institute of Physics, Jagiellonian University, 
ulica \L{}ojasiewicza 11, PL-30-348 Krak\'ow, Poland}

\date{\today}

\begin{abstract}

We discuss electronic properties and their evolution for the linear chain of $H_2$ molecules in the presence of a uniform external force $f$ acting along the chain. The system is described by an extended Hubbard model within a \emph{fully microscopic approach}. Explicitly, the microscopic parameters describing the intra- and inter-site Coulomb interactions are determined together with the hopping integrals by optimizing the system ground state energy and the single-particle wave functions in the correlated state. The many-body wave function is taken in the Jastrow form and the Variational  Monte-Carlo (VMC) method is used in combination with an \emph{ab initio} approach to determine the energy. Both the effective Bohr radii of the renormalized single-particle wave functions and the many-body wave function parameters are determined for each $f$. Hence, the evolution of the system can be analyzed in detail as a function of the equilibrium intermolecular distance, which in turn is determined for each $f$ value. The transition to the atomic state, including the Peierls distortion stability, can thus be studied in a systematic manner, particularly near the threshold of the dissociation of the molecular into atomic chain. The computational  reliability of VMC approach is also estimated.

\end{abstract}

\maketitle


\section{\label{sec:level1}Introduction}
Theoretical  description of electronic systems, demanding a consistent incorporation of inter-electronic correlations, is one of the most  challenging tasks of condensed matter physics. At least two complementary strategies for finding a proper description of these complex systems are usually considered: \emph{ab-initio} oriented techniques and parametrized model approaches. The former refers primarily  to the application of quantum-chemical methods, for instance the \emph{Density Functional Theory} based techniques (e.g., DFT+U, LDA+DMFT), exact diagonalization (ED), \emph{post}-Hartree- Fock methods such as the Configuration Interaction (CI), Møller-Plesset perturbation theory etc., applied to particular physical or chemical systems. The latter approaches, use e.g. the Hubbard  ~\cite{Hubbard} or $t-J$ ~\cite{Zegrodnik} models and their variants to encompass the  \emph{essential features} of electronic correlations  such as an unconventional superconductivity observed in  the cuprates ~\cite{Zegrodnik,Zegrodnik2,Zegrodnik3}, or the \emph{Mott-Hubbard} ~\cite{Hubbard,Gebhard}  transition in the transition metal oxides. Another example is the problem of the  the  solid (molecular) hydrogen metallization at extreme pressure ~\cite{Wigner}. In fact, the last issue comprises the most of challenges which are characteristic for both of the above mentioned methods \cite{Azadi,McMinis,Drummond}. It is believed that the metallization may occur by means of a transformation from the molecular crystal into the atomic one, i.e., $H_2$ molecules dissociate in atomic structure, wchich become metallic  at a critical pressure. This transition was first proposed by Wigner and Huntington ~\cite{Wigner} and is still under an intensive debate ~\cite{Castelvecchi,Liueaan}. While the phase diagram of the solid hydrogen is surprisingly complex ~\cite{Dias,Howie,Dalladay-Simpson} and predicted phase boundaries strongly depend on subtle effects such as a precise inclusion of the lattice dynamics ~\cite{Borinaga,Azadi2,Azadi}, the simplified models may provide insight into the electronic properties in  vicinity of the pressure-induced  molecular to atomic crystal transformation. 
As an illustration of this, we may quote the Mott-Hubbard-like transitions proposed by us recently in the low-dimensional hydrogenic systems \cite{Kadzielawa,Biborski2}. \\
\indent Previously we have used the \emph{Exact Diagonalization + Ab Initio} (EDABI) method \cite{Spalek,Zahorbenski,Kadzielawa2,Kadzielawa3,Biborski1,RycerzPhD} and could handle  only a relatively small number, typically up to $N<16$ ~\cite{RycerzPhD,Biborski1,Giner} atoms.
Therefore,  we have decided to replace here the \emph{exact} diagonalization  of the Hamiltonian matrix  by means of a \emph{Variational  Monte-Carlo} (VMC)  solution ~\cite{Becca,Rugers,Foulkes}. This allows us to  analyze the model of molecular hydrogen chain consisting of dozens of atoms, which can serve as an extension of the well known computational "\emph{benchmark}" results for an equally spaced chain composed of  hydrogen atoms ~\cite{Stella,Motta,RycerzNJP}.
According to Peierls theorem~\cite{Peierls}, such  a chain for one electron per atom, i.e., at the \emph{half-filling}, is unstable against spontaneous alternating distortion. However, this statement can be proved rigorously only in the absence of  electron-electron correlations ~\cite{Ovchinnikov}. The energetical stability of the correlated and distored chain were performed both for  the parametrized models   (c.f. e.g., ~\cite{Ovchinnikov,Hirsch}), as well as in the paradigm of \emph{ab-initio} method, see e.g. \cite{Motta,Stella,Giner}. The Peierls dimerization in $H_n$ rings and (finite) chains within Full-CI formalism with the maximal $n=14$ and open boundary conditions  were studied by Giner et al. ~\cite{Giner}. The linear hydrogen chain and its metallic properties in the framework of VMC method were analyzed by Stella et al.~\cite{Stella}, the state of art methods regarding this topic were also presented recently by Motta at al. ~\cite{Motta}. The dimerized $H_2$ chains but for the limited range of the lattice spacing were also investigated in the framework of the Diffusion Monte-Carlo method (DMC) ~\cite{Umari}. \\
\indent Here we follow a different approach, in which we start from the molecular-$H_2$ chain and stability of which is tuned by an external force.   We thus provide methodology  analogous to our EDABI-based studies \cite{Spalek,Zahorbenski,Kadzielawa2,Kadzielawa3,Biborski1,RycerzPhD}.
Applying an axial force (generalized pressure) to the system  -- which is a sole \emph{control} parameter -- we are able to construct the phase diagram and analyze electronic properties of the chain in the molecular (low-pressure) and nearly atomic (high-pressure) regimes. Namely, we study the chain distortion  as a function of pressure and discuss the role of the system size via the finite-size scaling procedure. \\
\indent In the following  sections we describe the model, its parametrization and provide computational details (cf. Sec. ~\ref{sec:ModelMethod}). Next, we analyze the phase diagram and electronic properties of the system from perspective of the force-induced dissociation into an atomic phase. We also analyze explicitly the effect of system size and its role in the  dissociation process by performing the finite-size scaling in Sec.~\ref{sec:Results}. We conclude and list further  issues to be scrutinized next.

\section{Model and Method}
\label{sec:ModelMethod}

\subsection{Molecular chain}
\begin{figure}
 \includegraphics[width=\linewidth]{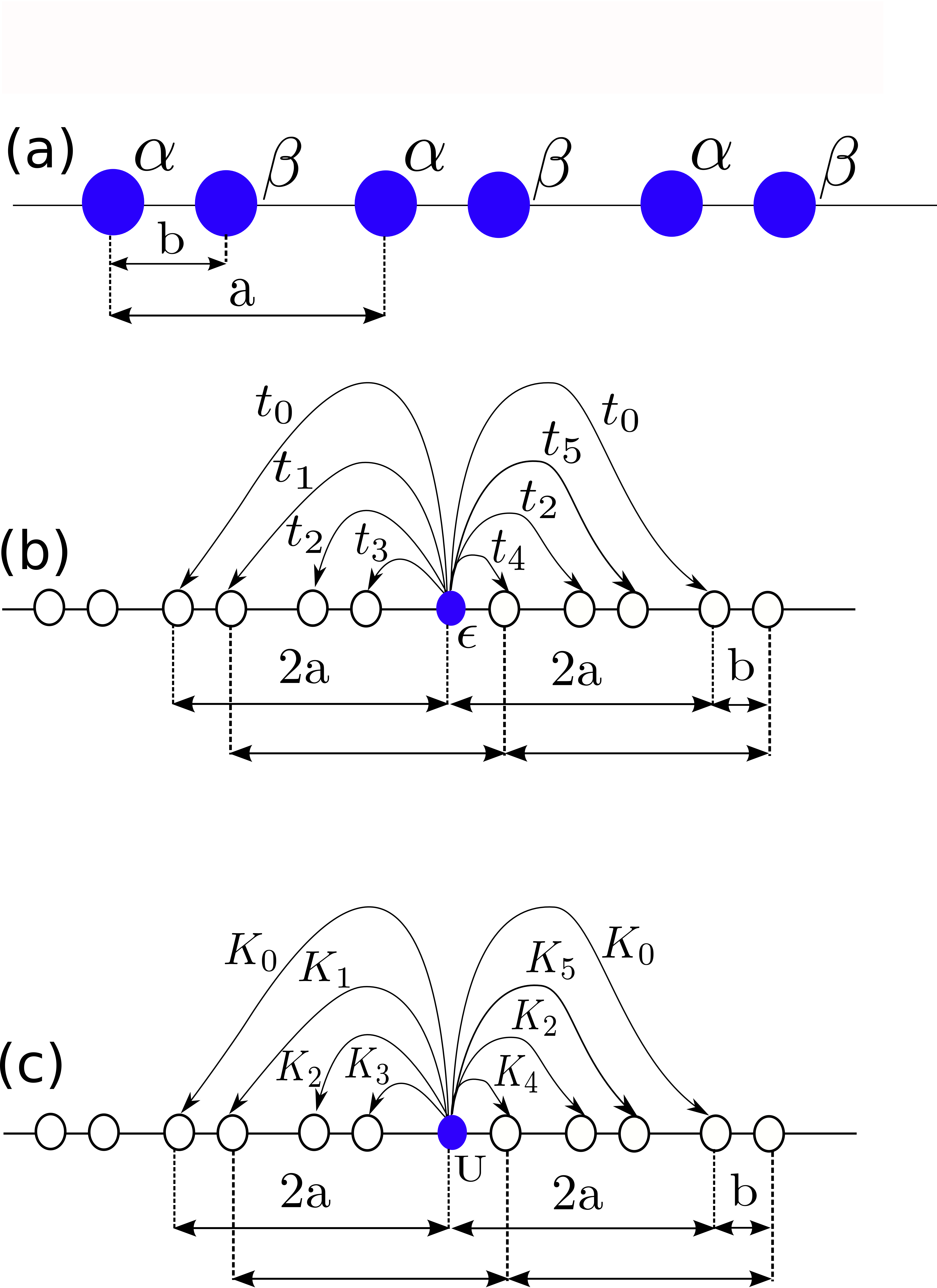}
 \caption{(a) Schematic representation of molecular chain, characterized by lattice parameter $a$ and  bond length $b$; the atomic center of  molecule are labelled as $\alpha$ or $\beta$; (b) hopping terms range extends up to $2a$. Note that only one atom $\alpha$ (blue circle) is marked for the sake of  clarity. However, by symmetry the same hopping configuration holds for $\beta$-centers; (c) Same as in (b), but for the intersite ($K$) interactions.} 
 \label{fig:model}
\end{figure}

We consider a linear hydrogenic-like molecular  chain (MLC) characterized by intermolecular distance (lattice parameter) $a$ and bond length $b$  (cf. Fig.~\ref{fig:model}). Note that for $b=a/2$ the system reduces to an \emph{atomic linear chain} (ALC). While each molecule consists of two atomic centers assigned as $\alpha$ and $\beta$; the corresponding Wannier functions are $w_{i,\alpha}(\vec{r})$ and $w_{i,\beta}(\vec{r})$  for the  $i$-th molecule. Orbitals $w_{i,\mu}(\vec{r})$,where $\mu=\{\alpha,\beta\}$, are assumed to be \emph{finite} contractions of $1s$ Slater atomic orbitals 

\begin{align}
\label{eq:slater}
	\psi_{i}^{\mu} \left(\vec{r}\right) \equiv \sqrt{\frac{\zeta^3}{\pi}} e^{-\zeta\left| \vec{r} - \vec{R}^{\mu}_{i} \right| },
\end{align}

where $\zeta$ may play the role of a variational parameter and $\vec{R}^{\mu}_{i}$ is its atomic position. In that situation

\begin{align}
 \label{eq:wannier_definition}
 w_{i,\mu} (\vec{r}) \approx \sum_{j(i,\nu)}^{L(i,\mu)}\sum_{\nu \in \{\alpha,\beta\}} c_{j\nu} {\psi}^{\nu}_{j} \left(\vec{r} \right),
\end{align}
with $L(i,\nu)$ and $j(i,\nu)$  are specific functions mapping the indices to the assumed \emph{cut-off radius} $r_{f}=3a$  in the \emph{tight binding} approximation. Additionally, we impose the orthogonality of $\{w_{i,\mu}(\vec{r})\}$ basis, i.e.,

\begin{align}
 \label{eq:wannier_orthogonality}
 \bracket{w_{i,\mu} (\vec{r})}{w_{j,\nu} (\vec{r})}=\delta_{\mu\nu}\delta_{ij},
\end{align} 
which in practice is ensured in terms of performing the L\"owdin symmetric orthogonalization for a block of molecules of size  exceeding the interactions range (see next Section).  The expansion coefficients  $c_{j\nu}$ are taken for both atoms forming the central molecule in a block  and the resulting Wannier functions $w_{i,\mu} (\vec{r})$ are repeated periodically. This procedure allows to assure their mutual orthogonality within desired accuracy.

\subsection{Hamiltonian and microscopic parameters}

As in our previously related ~\cite{Biborski1,Biborski2,Kadzielawa,Kadzielawa2}, we assume that Hamiltonian is of the  extended Hubbard form, i.e.,

\begin{align}
 \label{eq:hamiltonian_tot}
 \hat{\mathcal{H}} =& \sum_{i\mu} \epsilon_i^{\mu} \NUM{i}{0} + \sideset{}{'}\sum_{ij\mu\nu\sigma} t_{ij}^{\mu\nu} \CR{i\mu}{2}\AN{j\nu}{2} \\\notag
 & + U \sum_{i,\mu} \NUM{i\mu}{1} \NUM{i\mu}{-1} + \frac{1}{2}\sideset{}{'}\sum_{ij\mu\nu} K_{ij}^{\mu\nu} \NUM{i\mu}{0} \NUM{j\nu}{0} \\\notag 
 & + \frac{1}{2}\sum_{ij}\frac{2}{|\vec{R_{i}} -\vec{R_{j}}|},
\end{align}

where $\CR{i\mu}{2}(\AN{i\mu}{2})$ is fermionic creation (anihilation) operator, the local particle number  operator is $\NUM{i\mu}{2}\equiv\CR{i\mu}{2}\AN{i\mu}{2}$ and counts electrons of spin $\sigma$ at lattice site $i$ and for atom labelled bu $\mu=\alpha,\beta$. We also define total number of electrons  $\NUM{i\mu}{0}\equiv\NUM{i\mu}{-1}+\NUM{i\mu}{1}$. The primed summations emphasizes the exclusion of cases related to  $i=j \wedge \mu=\nu$. One--electron  matrix elements: atomic energy $\epsilon^{\mu}_{i}\equiv t_{ii}^{\mu\mu}$ and hopping amplitudes $t_{ij}^{\mu\nu}$, are defined (in the atomic units) so that
\begin{align}
 \label{eq:one-body}
 t_{ij}^{\mu\nu}  \equiv \matrixel{w_{i,\mu}(\vec{r})}{ -\nabla^{2} - \sum_{l=1}^{N_S}\frac{2}{|\vec{R_l} - \vec{r}|}}{w_{j,\nu}(\vec{r})},
 \end{align}
where $N_{S}$ is number of neighbors in \emph{interaction cut-off} sphere characterized by radius $r_{in}$. The intra-site $U$ and inter-site $K_{ij}^{\mu\nu}$ parameters  are the special cases of the general form of the interaction matrix elements
\begin{align}
\label{eq:microscopicP}
  V_{ijkl}^{\mu\nu\tau\rho} \equiv \matrixel{w_{i\mu}(\vec{r}) w_{j\nu}(\vec{r'})}{\frac{2}{|\vec{r} - \vec{r'}|}}{w_{k\tau}(\vec{r})w_{l\rho}(\vec{r}')},
\end{align}
i.e., $U=V_{iiii}^{\mu\mu\mu\mu}=V_{iiii}^{\nu\nu\nu\nu}$ and $K=V_{ijij}^{\mu\nu}$.
We ensure that all integrals are well defined by means of  assumption that $r_{in}=2a<R_{cf}$, i.e., Eq. \eqref{eq:wannier_definition} is always fulfilled. For the sake of brevity we number all the considered hoppings $t_{ij}^{\mu\nu}$ and interaction parameters $K_{ij}^{\mu\nu}$, as in Fig.~\ref{fig:model}. According to the fact that single-electron wave functions $w_{i,\mu} (\vec{r})$ are real and taking into account system symmetries, selected  hopping and interaction parameters are identical. \\ The last term in the Hamiltonian \eqref{eq:hamiltonian_tot} describes Coulomb interactions between ions which we treat in a classical manner. We neglect lattice dynamics and \emph{electron-phonon} coupling, which is in principle possible to be included in the VMC scheme ~\cite{Watanabe,Ohgoe}. In this context, its inclusion would complicate excessively  our computational procedure.

\subsection{Variational  Monte-Carlo}
We employ VMC method for finding an  approximate ground state of the system described by Hamiltonian \eqref{eq:hamiltonian_tot}. As a \emph{variational ansatz} for $N$-electron wave function  we  choose the \emph{trial state} $\ket{\Psi_T^{N}}$ of the form

\begin{align}
\label{eq:ansatz}
    \ket{\Psi_T^{N}}\equiv\hat{\mathcal{P}}\ket{\Phi_{FE}},
\end{align}
where $\mathcal{P}$ is the Jastrow factor 
\label{eq:jastrow}
\begin{align}
    \hat{\mathcal{P}}=\text{exp}\Bigg[-\sideset{}{'}\sum_{i\mu,j\nu}\lambda_{i\mu,j\nu}\NUM{i\mu}{0}\NUM{j\nu}{0}-\sum_{i\mu}\lambda_{i\mu}\NUM{i\mu}{1}\NUM{i\mu}{-1}\Bigg],
\end{align}
 specified by the set of variational parameters $\{\lambda_{i\mu j\nu},\lambda_{i\mu}\}$ and provides a sufficient flexibility to include electronic correlations, while  $\ket{\Phi_{FE}}$ is the solution for the system of non-interacting electrons, i.e., for the case  $U=K_{ij}^{\mu\nu}=0$. 
The uncorrelated  solution  $\ket{\Phi_{FE}}$ may be written as an expansion in the   basis $\{\ket{\text{x}}\}$  spanning  $N$-electron Fock space, i.e.,
\label{eq:phi}
\begin{align}
   \ket{\Phi_{FE}} = \sum_{x}c_{x}{\ket{\text{x}}},
\end{align}
with
\label{eq:x_state}
\begin{align}
   \ket{\text{x}} \equiv \ket{\text{x}_{\uparrow}}\otimes\ket{\text{x}_{\downarrow}}=\prod_{i\mu(\ket{\text{x}_{\uparrow}})}^{n_{\uparrow}(\ket{\text{x}_{\uparrow}})}\CR{i\mu}{1}\prod_{j\nu(\ket{\text{x}_{\downarrow}})}^{n_{\downarrow}(\ket{\text{x}_{\downarrow}})}\CR{j\nu}{-1}\ket{\text{0}},
\end{align}
where  $i\mu,j\nu$ are single particle state indices and the total number of spin-up and spin-down electrons ($n_\uparrow$,$n_\downarrow$ respectively); are mapped from the spin-configuration sectors $\ket{\text{x}_\uparrow}$ and $\ket{\text{x}_\downarrow}$, with $\ket{0}$ being the vacuum state. The average of an operator $\hat{\mathcal{O}}$ is given as
\label{eq:observable_avg}
\begin{align}
   \langle \hat{\mathcal{O}} \rangle \equiv \frac{\matrixel{\Psi_T^{N}}{\hat{\mathcal{O}}}{\Psi_T^{N}}}{\bracket{\Psi_T^{N}}{\Psi_T^{N}}} = \frac{\sum_{x}\bracket{\Psi_T^{N}}{\text{x}}\matrixel{\text{x}}{\hat{\mathcal{O}}}{\Psi_T^{N}}}{\sum_{x}\bracket{\Psi_T^{N}}{\text{x}}\bracket{\text{x}}{\Psi_T^{N}}},
\end{align}
and may be expressed in terms of its \emph{local value} $O_{loc}(x)$
\label{eq:observable_local_avg}
\begin{align}
   \langle \hat{\mathcal{O}} \rangle =  \sum_{x}\rho(x)O_{loc}(x),
\end{align}
where 
\label{eq:observable_local_avg_approx}
\begin{align}
   O_{loc}(x) \equiv \frac{\matrixel{\text{x}}{\hat{\mathcal{O}}}{\Psi_T^{N}}}{\bracket{\text{x}}{\Psi_T^{N}}}
\end{align}
and 
\label{eq:probability_density}
\begin{align}
\rho(x)\equiv \frac{\bracket{\Psi_T^{N}}{\text{x}}\bracket{\text{x}}{\Psi_T^{N}}}{\sum_{\text{x'}}\bracket{\Psi_T^{N}}{\text{x'}}\bracket{\text{x'}}{\Psi_T^{N}}} 
\end{align}
is regarded  as the \emph{probability density function}.
Eventually, sampling $M$ states $\ket{\text{x}}$ from the distribution governed by $\rho(x)$ - in our case performed in a standard manner i.e., by  means of application of Metropolis algorithm - allows to obtain an approximate value  $\langle  \hat{\mathcal{O}}\rangle$ in the form
\begin{align}
\label{eq:vmc_essence}
\langle\hat{\mathcal{O}}\rangle \approx \frac{1}{M}\sum_{m=1}^{m=M}O_{loc}(x_m).
\end{align}
In particular, the total trial energy is
\begin{align}
\label{eq:vmc_energy}
E_T(\{\lambda_{i\mu,j\nu},\lambda_{i\mu}\},\zeta) = \langle\hat{\mathcal{H}}\rangle,
\end{align}
and its variance
\begin{align}
\label{eq:vmc_energy_variance}
\delta E_T^2(\{\lambda_{i\mu,j\nu},\lambda_{i\mu}\},\zeta) \equiv \frac{1}{M}\sum_{x}\Big[\langle\hat{\mathcal{H}}\rangle - H_{loc}(x)\Big]^{2}
\end{align}

can be computed. The trial energy, its variance or a linear combination of both may be used for the optimization leading to an approximate ground  state. 

\subsection{Numerical procedure}

We analyze the possibility  of the molecular-chain dissociation into the atomic, possibly metallic state, within a fully microscopic approach developed and tested by us earlier. The external, collinear force $f$, is regarded here as an applied pressure  on this translationally invariant $1D$ system. This  is the sole factor, which  controls such an atomization. For simplicity, we also impose the periodic boundary conditions (PBC) to eliminate the boundary effects. The proper thermodynamic potential in this case is the enthalpy \cite{Kadzielawa,Biborski2} which takes  the form
\begin{align}
    \frac{H}{N}=h\equiv f\frac{a}{2}+\frac{E}{N},
\end{align}
where $h$ is the enthalpy per particle  and $E$ is the system internal energy. The equilibrium value of $h$ and the structural parameters $a$ and $b$ as well as $\zeta$ at given $f$ are all found by means of minimization of functional
\begin{align}
h(f;a,b,\zeta)=f\frac{a}{2} + \frac{E(a,b,\zeta)}{N}.
\end{align}

\begin{figure}
 \includegraphics[width=0.8\linewidth]{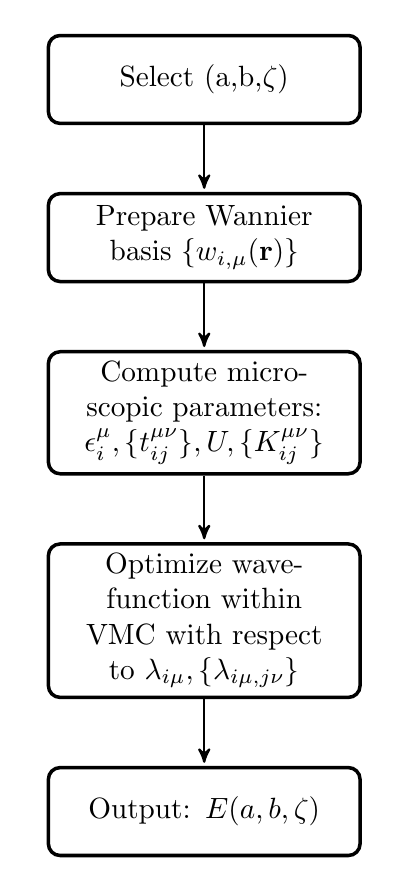}
 \caption{Energy calculation flow chart. } 
 \label{fig:flow_chart}
\end{figure}

In practice to perform optimization one must be able to compute the $E(a,b,\zeta)$ (cf. Fig. ~\ref{fig:flow_chart}).  The optimization run of VMC is carried out by minimiazing of $\delta E_T^2$ defined by  Eq.\eqref{eq:vmc_energy_variance}. Note, that apart from the optimization of the Jastrow parameters, the energy must be minimized with respect to  $\zeta$. The energy  which is optimized with respect to $\zeta$ in a given range of $a$ and $b$ allows in turn  find the minima of $h(a,b)$. 
The main computational effort  is   optimization   with respect to $\zeta$  for each considered pair $a,b$.  The algorithm consist of: (\emph{i}) single-particle basis orthognalization \ref{eq:wannier_orthogonality}, (\emph{ii}) computation of microscopic parameters according to  Eqs.\eqref{eq:one-body} and \eqref{eq:microscopicP}; (\emph{iii}) optimization with respect to Jastrow variational parameters and $\zeta$. We assume the  \emph{cut-off radius} for Jastrow factor parameters as $r_{P}=r_{in}=2a$, and therefore,  the  number of interaction parameters, hoppings (with on-site atomic energy present) and the Jastrow variational parameters are equal which amounts to seven independent quantities. The results of calculations presented in the following sections are obtained by means of  self-developed codes, available from our computational library \textbf{Q}uantum \textbf{M}etallization \textbf{T}ools (QMT)~\cite{qmtURL}.

\section{Results} 
\label{sec:Results}
\subsection{Reliability of results}
While the quality of results obtained by means of utilization of  chosen wave function \emph{ansatz} (Eq. \eqref{eq:ansatz}) is not \emph{a priori} known, we have performed the testing calculations for $N=10$, i.e., number of particles  which is attainable by our \emph{exact} treatment ~\cite{Zahorbenski,RycerzPhD, Spalek,Kadzielawa,Kadzielawa2,Kadzielawa3,Biborski1,Biborski2}. Precisely, we have applied the EDABI method to inspect the validity of the data obtained by means of VMC.  
\begin{figure}
 \includegraphics[width=\linewidth]{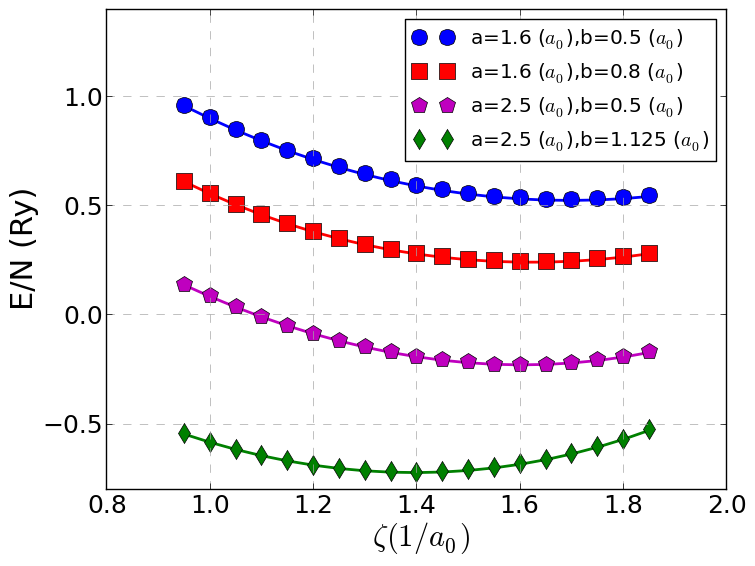}
 \caption{Energy per electron for the four different chain configurations as a function of variational parameter $\zeta$ for $N=10$. The symbols refer to the data obtained by means of VMC and the solid lines are  result of the exact diagonalization (EDABI). The size of the symbols is larger than an estimated statistical error, i.e., $\sim 10^{-3}$ (Ry) per electron. } 
 \label{fig:vmc_vs_exact_zeta}
\end{figure}
As one may deduce from Fig.~\ref{fig:vmc_vs_exact_zeta}, where the total system energy per electron versus $\zeta$ is plotted, the agreement between the exact and VMC results is very good; typical differences do not exceed the statistical error. The energy of the system  depends on $\zeta$, in some cases, e.g., for $a=1.6a_0$ and $b=0.5a_0$ (where $a_0$ is Bohr radius), is reduced even factor of two when compared to the non-renormalized case (i.e., for $\zeta=a_0^{-1})$. This observation confirms that though it increases computational complexity, the optimization with respect to $\zeta$ is  important, if not indispensable.
An additional remark is in place. According to $a/2=b$ case, one finds that by treating ALC strictly  as a specific variant of MLC reduce  the number of microscopic parameters as relative  to the MLC (e.g. $t_1=t_5$). However, this implies also a similar reduction of the number of $\lambda_{ij}$ parameters. The number of independent $\lambda_{i\mu j\nu}$ can be increased by   extending of the correlation radius, but in such a scenario ALC would be solved for the different form of the many-body wave-function ansatz. Therefore, we have  decided to introduce a small distortion $\delta b=10^{-5}$ to $a/2=b$ situation, i.e., $b\rightarrow b-\delta b$. In this manner we treat ALC as a nearly undistorted MLC to conform the consistency of the phase diagram. Results for  $a/2=b$ cases and presented in  the following Subsections mean that, strictly speaking $a/2 \approx  b- \delta b$.
\subsection{Phase diagram of finite system}
\begin{figure}
 \includegraphics[width=\linewidth]{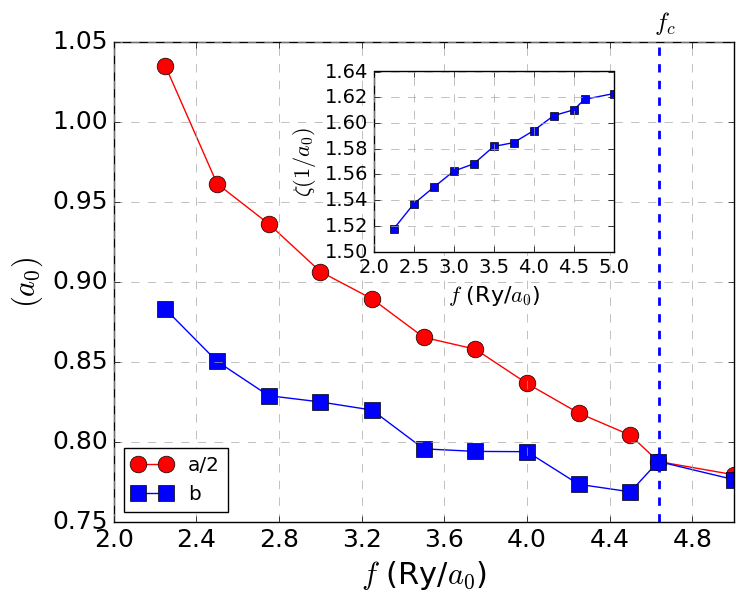}
 \caption{Equilibrium structural parameters $a$ and $b$ as a function of external force $f$ for $N=50$. Note that we plot $a/2$ instead of $a$ to help the identification of MLC $\rightarrow$  ALC  transition --  $a/2\rightarrow b$. The critical force value is $f_c\approx 4.64Ry/a_0$. The inset contains corresponding dependency $\zeta(f)$}. 
 \label{fig:aR50}
\end{figure}

We have performed the set of calculations for the systems consisting of $N=10,18,26,34,42,50,66$ and $74$ electrons with corresponding number of ions,  with imposed PBC. In a single VMC run, it is desired to ensure that the number of considered electrons allows for formation of  \emph{closed-shell} ~\cite{Becca} configuration within the trial wave function. We found out that the selection of  $N=34,42,50,66$ and $74$ meets this requirement; it is not the case for $N=10,18$, and $26$. Despite this shortcoming, the results obtained for  $N=10$ seem to be reasonable as might be deduced from Fig. ~\ref{fig:vmc_vs_exact_zeta} where we compare the robustness of our approach to the exact treatment. However, for the finite size scaling analysis we utilize data obtained for the five of  largest considered systems, i.e., for  $N=34,42,50,66$, and $74$.
\begin{figure}
 \includegraphics[width=\linewidth]{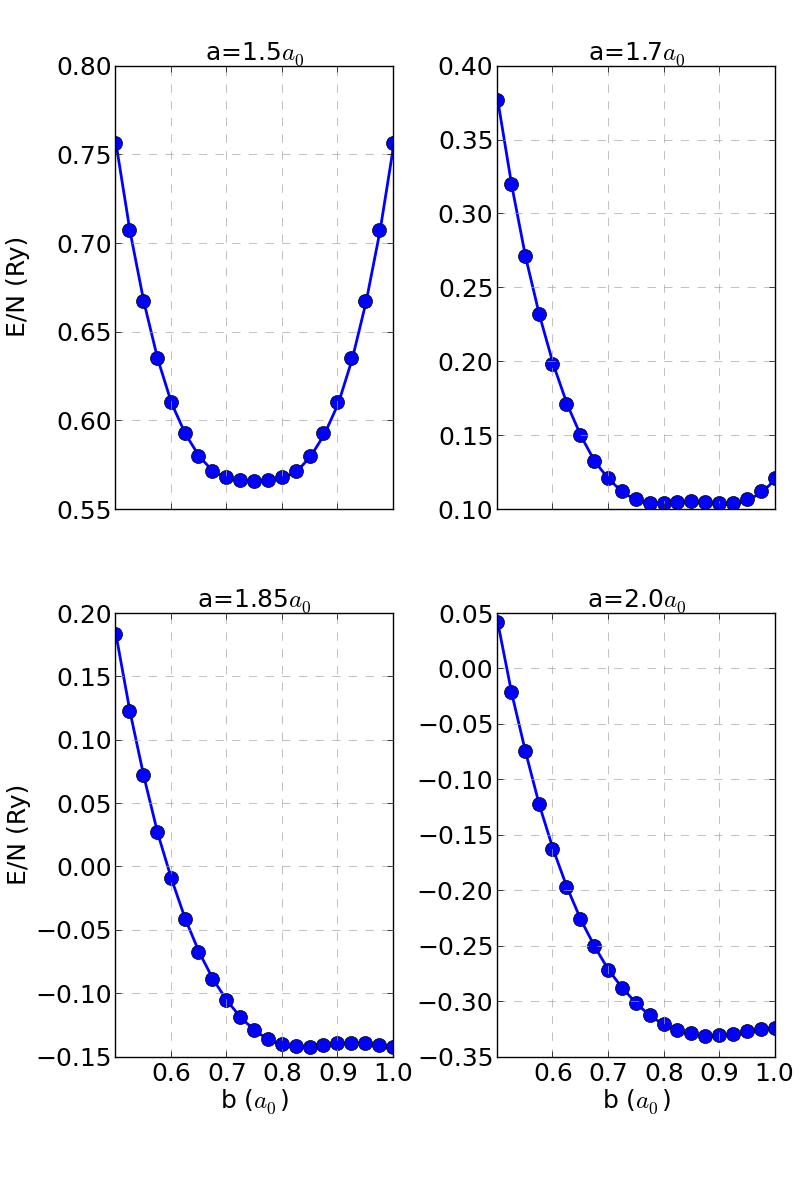}
 \caption{Total energy as a function of $b$ for selected $a=1.5a_0,1.7a_0,1.85a_0$ and $2.0a_0$ obtained by means of QVMC for $N=50$. At  fixed $a=1.5a_0$ system tends to ALC solution,i.e., the energy minimum is for $b=a/2$.} 
 \label{fig:iso_a}
\end{figure}
For each $N$ we have scanned the $(a,b)$ space to find the pressure (force) at which system undergoes a transition to the atomic state in question. The size and range of the mesh was varied with $N$ as the energy landscape depends on it (see next Subsection). However, we assumed a constant resolution, $\Delta a = \Delta b = 0.025a_0$. For the sake of clarity, in this Subsection we present results obtained for $N=50$ which are representative for the whole set, whereas, the important conclusions obtained from the \emph{finite-size-scaling} analysis are discussed in the following Subsection. Note the that results  for $N=66$ and $N=74$ have been obtained for the sub-range of $(a,b)$ if compared to those for $N\leq 50$ due to the computational time limitations. However, we can still to use these data to perform the finite-size scaling analysis.
\begin{figure}
 \includegraphics[width=\linewidth]{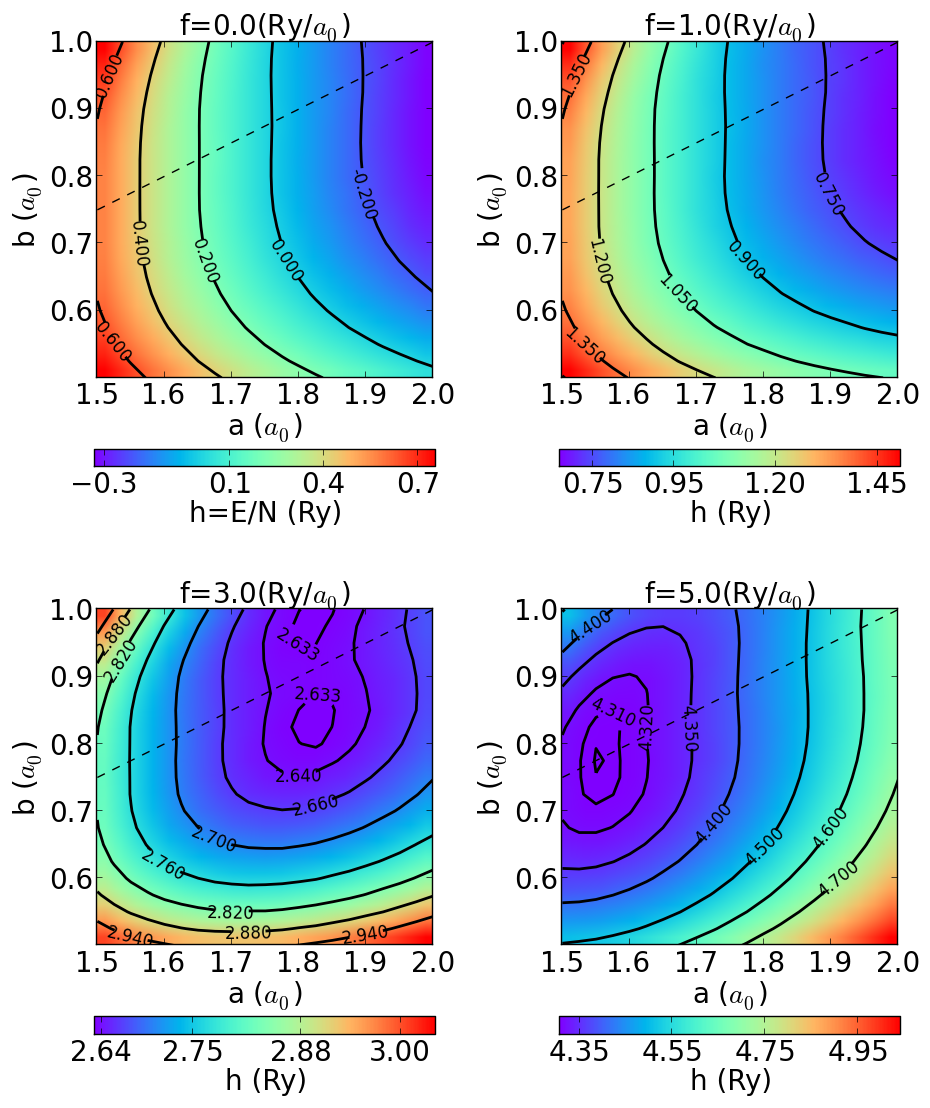}
 \caption{Enthalpy as a function of $a$ and $b$ for the selected values of $f$ with $N=50$. The dashed line indicates ALC. Plots are obtained from the mesh with resolution $\Delta a=\Delta b= 0.025 a_0$ by smoothing it within bilinear interpolation for the sake of clarity.} 
 \label{fig:phase_diagram_panel}
\end{figure}
In Fig.~\ref{fig:iso_a} we show the total energy as a function of $b$ for the selected $a$ isolines. With the increasing $a=1.7a_0,1.85a_0$ and $2.0a_0$, the two symmetric (for $b<a/2$ and $b>a/2$) minima appear and indicate the molecular chain stability (at fixed $a$). Results referring to $b>a/2$ are those obtained for $b<a/2$ but reflected with respect to the line $b=a/2$, in  accordance with the system symmetry. 
The evolution  of the location of the  minimum of enthalpy is reflected in the relation between  $a$ and $b$ (Fig.~\ref{fig:aR50}). 
As the force value is below  $f_c$, the system persists in molecular state, i.e., $a/2 > b$ whereas at $f_c$ it becomes atomic. We observe that while the \emph{bond-length} $a$ decreases monotonically, it is not the case for $b$. In  vicinity of the critical force, but for  $f<f_c$, the value of  $b$ increases  -- following the prior decrease  -- and finally attains the critical value $\approx a/2$. 
\subsection{Ground state Energy and the critical force: Finite size scaling }

\begin{figure}
 \includegraphics[width=\linewidth]{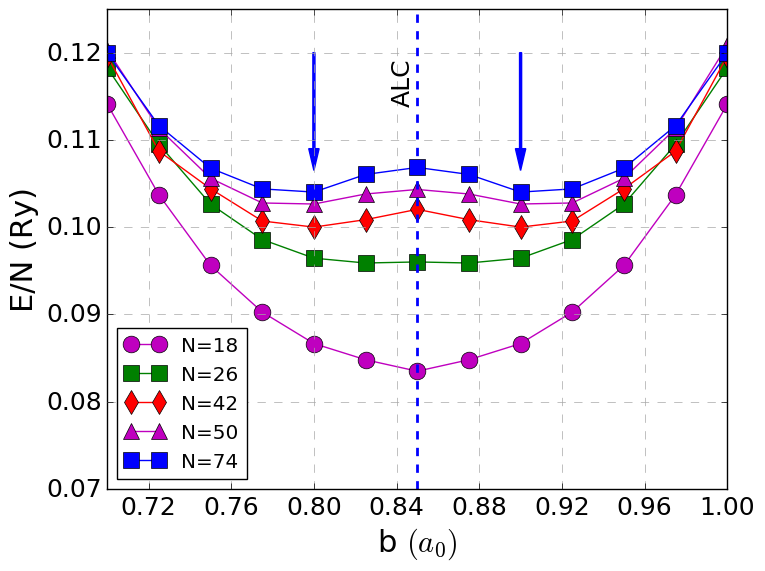}
 \caption{Energy per electron for fixed value of $a=1.7$ as function of $b$ for the selected $N$. Note the double minimum (marked by the arrows) appearing with the increasing system size. } 
 \label{fig:1.7a_scalin}
\end{figure}

As mentioned above, the results obtained for $N=50$  represent qualitatively the trend for other $N$ studied. Namely, we observe atomization of the chain for each considered system size. However, we find out  differences between them. In Fig.~\ref{fig:1.7a_scalin} we plot the energy as a function of $b$ for $a=1.7a_0$, for the specified values of $N$. The shape of the energy per atom vs $b$  for a fixed $a$ depends on the system size not only quantitatively, but also differs qualitatively. Namely  one sees (Fig.~\ref{fig:1.7a_scalin})  that the  $E$ minimum evolves from that corresponding to ALC for $N=18$ as to that for $N=26$, where  becomes flatter suggesting tendency towards the molecular solution. In effect, it  takes place  for $N\geq42$. The energy values do not differ substantially  for $b\simeq 0.7 a_0$, i.e., when the system is  deep in the  MLC  state, if compared to $b\simeq a/2$, where $b$  is in the  vicinity of that corresponding to the ALC solution. 

\begin{figure}
 \includegraphics[width=\linewidth]{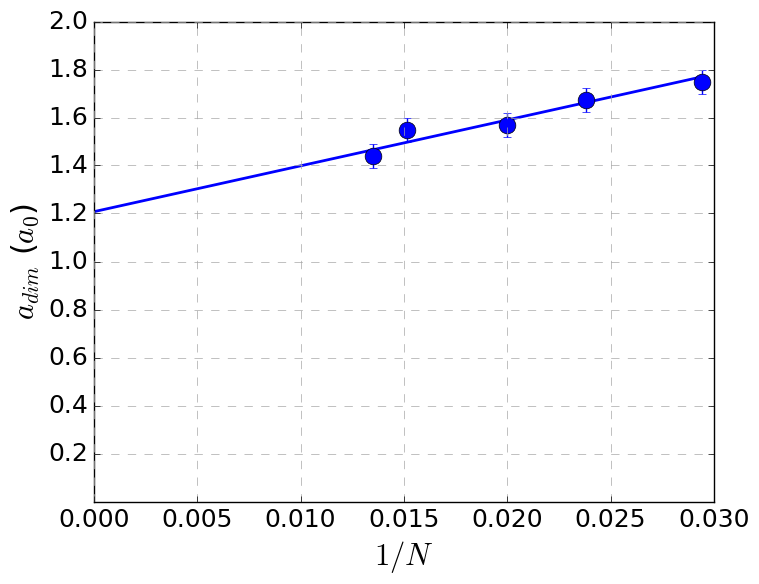}
 \caption{The relation between $a_{dim}$ and inverted system size $N$. The line indicates the linear fit.} 
 \label{fig:a_dim}
\end{figure}

\begin{figure}
 \includegraphics[width=\linewidth]{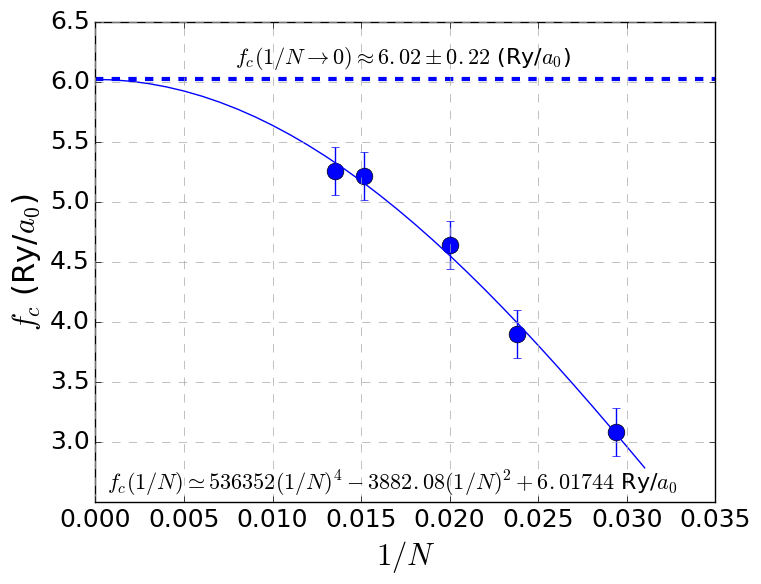}
 \caption{Critical force $f_c$ versus the inverse system size. Points with the error bars indicate obtained values. Solid line is a polynomial fit (explicitly written in a bottom)  whereas dashed line marks the estimated value of  $f_c(1/N\rightarrow 0)$. } 
 \label{fig:fc_scalin}
\end{figure}

We observe a \emph{shift} of the minimum  referring to the first occurrence of ALC on the $E(a,b)$ plane, i.e., to the highest possible $a=b/2\equiv a_{dim}$ as function of the system size. This behavior was also observed for the finite chains and rings  by Giner et al. \cite{Giner}.   Note that existence of $a_{dim}>0$ in the thermodynamic limit is a \emph{necessary} but not \emph{sufficient} condition for the suppression of the Peierls-like state. Therefore, we checked (within the accessible accuracy), if $a_{dim}(1/N\rightarrow 0) > 0$ in the energy landscape.  To answer the question if  dimerization is suppressed under certain value of the force in a thermodynamic limit, we analyzed  both $a_{dim}$ and $f_c$ as a functions of $1/N$ for $N=34,42,50,66$, and $74$,in the \emph{closed-shell} cases ~\cite{Becca}.  The $a_{dim}(1/N)$ for the considered $N$ exhibits a linear behavior (cf. Fig.~\ref{fig:a_dim}) so that $a_{dim}(1/N\rightarrow 0) \approx 1.77$.  Therefore, to answer the question if dimerization is suppressed in the thermodynamic limit at certain value of the applied force, we have performed finite size scaling of $f_c$ which is shown in Fig.~\ref{fig:fc_scalin}. We classify system as ALC at $f=f_c$, which corresponds to  abrupt decrease  of $a$ and  abrupt increase of $b$, such a $a/2 \approx b$ (cf. Fig.~\ref{fig:aR50}). In Fig.~\ref{fig:fc_scalin} we also plot $f_c(1/N)$ with the specified polynomial function fitted to data, finally obtaining $f_c(1/N\rightarrow 0)\approx 6.02 \pm 0.22 (\text{Ry}/a_0)$.  This value may seem to be overestimated, since tendency  for $N=66$ and $N=74$  is to suppress  of the distored (Peierls-like) state.\\
\indent For the sake of comparison  we have performed also calculations of the electronically non-interacting system (cf. Sec.~\ref{sec:Summary}. We observed that the distortion at $f=0$ (for which this approach predict absolute minimum of the enthalpy) is very small, i.e., $a-2b = 1.3988a_{0} - 2 \times 0.6919a_{0}\approx 0.015a_{0}$. Therefore, it may be concluded that in the regime of the lower range of $f$ (i.e., for $a/2 > b$), the correlations enhance the  distortion magnitude.

\section{Electronic properties}
 From the finite size scaling follows  that ALC is a stable configuration for \emph{finite} value $f$ in the thermodynamic limit. Therefore,  we provide next the basic electronic properties for both the MLC and ALC states, particularly in the regime $a/2\approx b$. 
\subsection{Charge energy gap}
\begin{figure}
 \includegraphics[width=\linewidth]{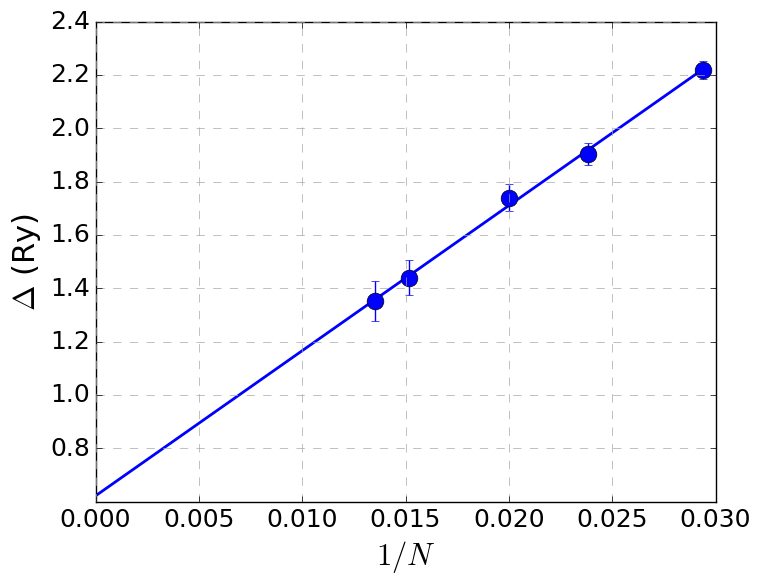}
 \caption{Exemplary finite size scaling of the charge gap $\Delta$ at $f=3 \text{Ry}/a_0$. The line represents the linear fit.} \label{fig:1.5cg}
\end{figure}

\begin{figure}
 \includegraphics[width=\linewidth]{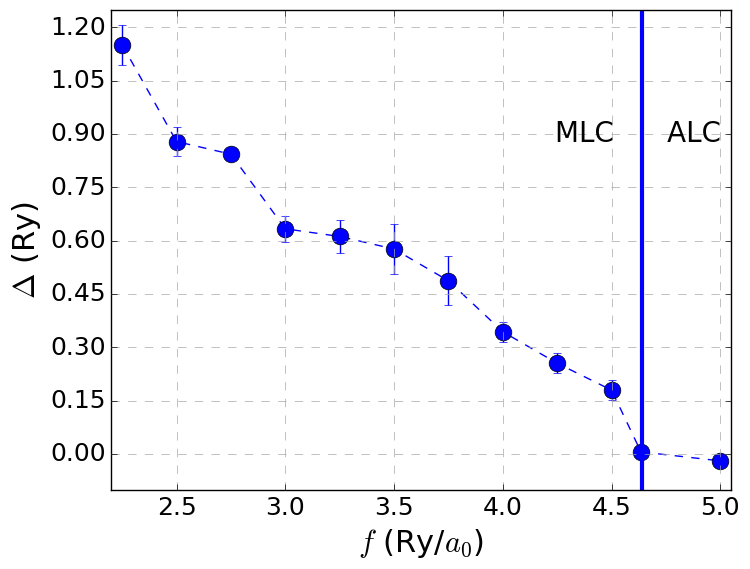}
 \caption{Extrapolated  charge gap $\Delta(f)$ for $a(f),b(f)$ referring to $N=50$ (see text). The vertical line separates MLC and ALC (for $N=50$), whereas the dashed line is as a guide to the eye.} \label{fig:charge_gap}
\end{figure}

\begin{figure}
 \includegraphics[width=\linewidth]{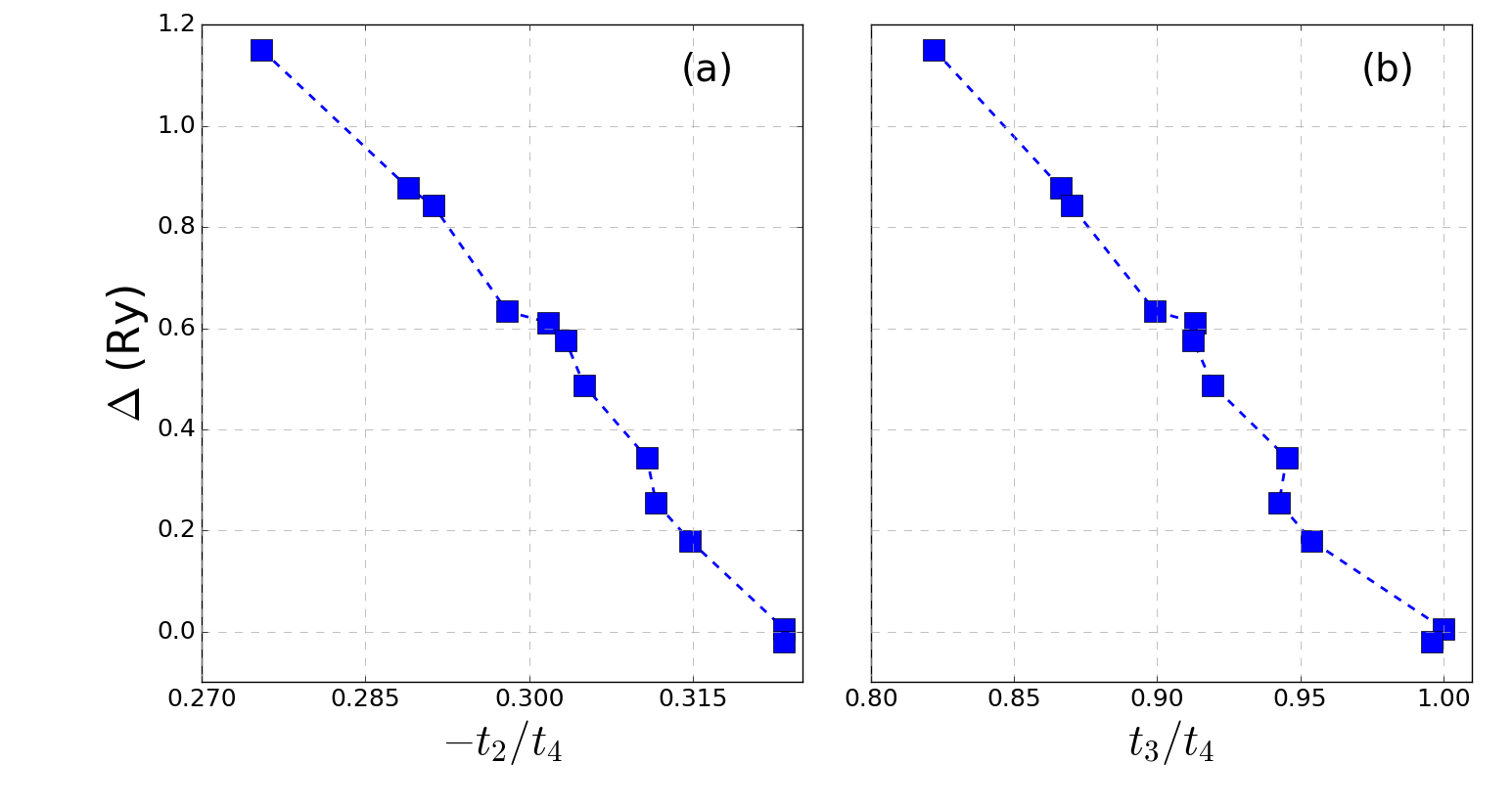}
 \caption{The charge energy gap as a function of  the selected ratios between hoppings. } \label{fig:hopps_gap}
\end{figure}

To provide an evidence for insulating or metallic character of MLC close to the ALC solution we estimated the \emph{charge energy gap} \cite{Ejima,Gebhard,Japaridze} 
\begin{align}
\label{eq:energy_gap}
\Delta \equiv \big[-2E(N)+E(N-4)+E(N+4)\big]/4.
\end{align}
This form of $\Delta$ allowed us to accomplish the closed-shell configuration, since for the considered system sizes (i.e., $N \in \{34,42,50,66,74\}$ we observe the four-fold degeneracy (including spin) in $\ket{\Phi_{FE}}$ for the highest occupied and the lowest unoccupied levels.
We intended to isolate the size $N$ as a single scaling parameter  and we were not able to perform the conclusive scaling for $a(1/N)$ and $b(1/N)$ due to the limited accuracy and maximal available value of $N$. Moreover, performing the scaling of  $\Delta(1/N)$ for  given  $f$  provides additional complication, namely at given $f$ the two systems of the different size, $N_1 < N_2$,  may correspond to ALC and MLC solutions, respectively, what in turn reduces the available number of points to be fitted, especially close to the ALC. On the other hand, we have intended to single out, at least qualitatively, the electronic characteristics of the molecular and atomic systems. Therefore,  we have computed $\Delta$ for $N \in \{34,42,50,66,74\}$ for configurations $a(f),b(f)$ referring to those obtained for $N=50$. It means that microscopic parameters (hoppings, interactions, ions repulsion) remained function of $f$ disregarding $N$. 
In Fig.~\ref{fig:1.5cg} we present exemplary $\Delta(1/N)$ dependence .  
We performed set of the linear fits (cf. Fig.~\ref{fig:1.5cg}) to obtain $\Delta(1/N \rightarrow 0)$ limit by means of extrapolation which eventually provided provided $\Delta(f)$ dependency (Fig.~\ref{fig:charge_gap}). The MLC system exhibits insulating characteristics as expected for the Peierls like state, however in the vicinity of ALC the gap seems to be small or vanishing. In the ALC state the gap is closed  indicating appearance of metallic state, in agreement with the full Hamiltonian solution obtained by Stella et al. \cite{Stella}. Indeed, for $f=5 \text{Ry}/a_0$ the ALC is stable with $a/2\approx b=0.776a_0$. This value refers to the range of $a/2$ where  hydrogenic atomic linear chain  is claimed to be metallic \cite {Stella}.
Note also, that sudden decrease of $\Delta$ to $\approx 0$ at $f_c$ \emph{coincides} with the claimed dissociation. 
The $\Delta$ has also a clear dependence on the hopping ratios $-t_2/t_4$ and $t_3/t_4$ (cf. Appendix A for all the values of microscopic parameters), as is shown in (Fig.~\ref{fig:hopps_gap}ab). The $t_2$ is positive and the ratio $-t_2/t_4$ increases with the increasing $f$. The charge gap closes at $\approx t_2/t_4\approx 0.32$, whereas  remains close to the value of $t_2$, reaching unity in the ALC limit, as expected. The charge gap closure with the increasing $-t_2/t_4$  resembles behavior observed in the $t-t'$ Hubbard  model, where the metallicity is induced by the increasing ratio between second ($t'$)  and the nearest ($t$) neighbor  hopping amplitudes \cite{Japaridze}, i.e., $-t'/t$.  The relative (to $t_4$) increase of the hopping amplitude, which we observe for $t_0,t_1,t_2$ and $t_3$   plays the role in the microscopic mechanism of the metallization.\\

\subsubsection{Correlation functions}
As in related studies \cite{Hohenadler,Wang,Sorella,RycerzPhD}   we consider next the  \emph{density-density}  and \emph{spin-spin} correlation functions to provide  evidence (if any) for the charge density and spin order in the system. We define the \emph{density-density} correlations via
\begin{align}
\label{eq:density_corel}
C_{i\mu,j\nu}\equiv\langle{\NUM{i\mu}{0}\NUM{j\nu}{0}} \rangle -\average{\NUM{i\mu}{0}}\average{\NUM{j\nu}{0}},
\end{align}
and, the \emph{spin-spin} correspondants
\begin{align}
\label{eq:spin_corel}
S_{i\mu,j\nu}\equiv\average{(\NUM{i\mu}{1}-\NUM{i\mu}{2})(\NUM{j\nu}{1}-\NUM{j\nu}{2})}= \langle{\hat{S}^{z}_{i\mu}\hat{S}^{z}_{j\nu}} \rangle.
\end{align}
\begin{figure}
 \includegraphics[width=\linewidth]{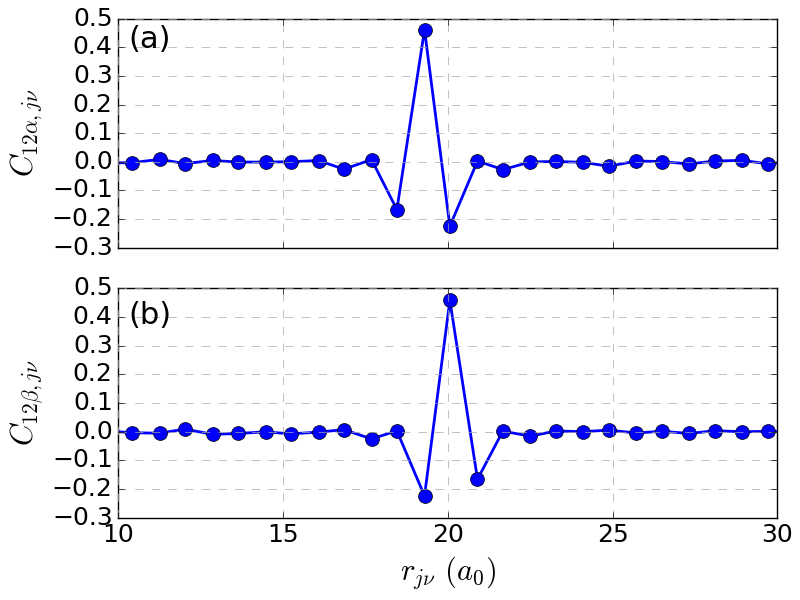}
 \caption{Density-density correlation functions plotted for the results obtained at $f=4.5 \text{ Ry}/a_0$ and $N=50$ for both $\alpha$ (a) and $\beta$ (b) sites, respectively.} 
 \label{fig:ccc}
\end{figure}
\begin{figure}
 \includegraphics[width=\linewidth]{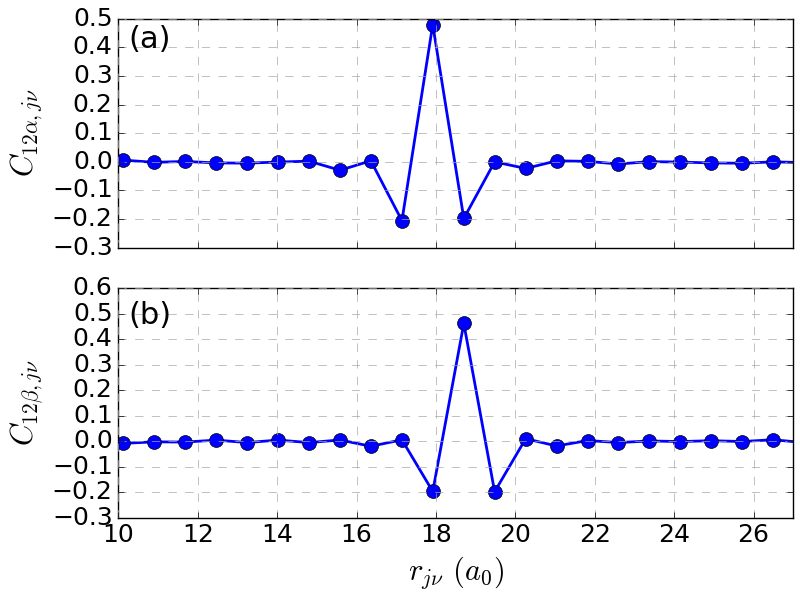}
 \caption{Density-density correlation functions plotted for the results obtained at $f=5\text{ Ry}/a_0$ and $N=50$ for both $\alpha$ (a) and $\beta$ (b) sites. Note that distinction between $\alpha$ and $\beta$ is arbitral since $a/2\approx b$. } 
 \label{fig:alc_ccc}
\end{figure}
In Fig. \ref{fig:ccc} we plot the density-density correlation functions for both $\alpha$ and $\beta$ sites. The oscillations of $C_{i\mu j\nu}$ decay at relatively  short distances. In fact, the amplitude exceeds the statistical noise only for the nearest and the next nearest neighbors which correspond to he distances  $r_{j\nu}=b$ and $r_{j\nu} = a-b$ respectively. While the system is still in MLC state at chosen $f$, damped oscillations are antisymmetric as $a > 2b$, contrary to those obtained for  ALC (cf. Fig. ~\ref{fig:alc_ccc}). Oscillations limited to the two nearest sites  suggest, that there is no   \emph{charge order}/\emph{charge density wave} (CDW) present in the system.
\begin{figure}
 \includegraphics[width=\linewidth]{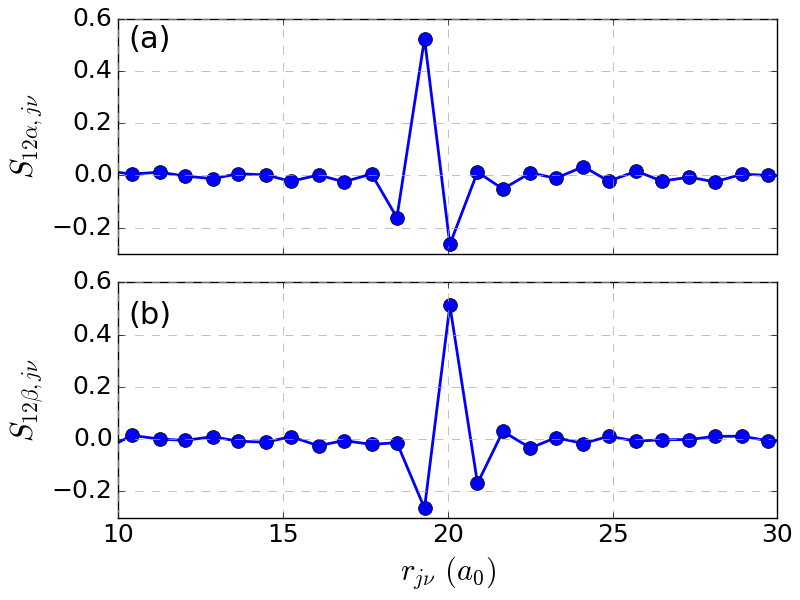}
 \caption{Spin-spin correlation functions plotted for the results obtained at $f=4.5\text{ Ry}/a_0$ and $N=50$ for both $\alpha$ (a) and $\beta$ (b) sites.} 
 \label{fig:sss}
\end{figure}
\begin{figure}
 \includegraphics[width=\linewidth]{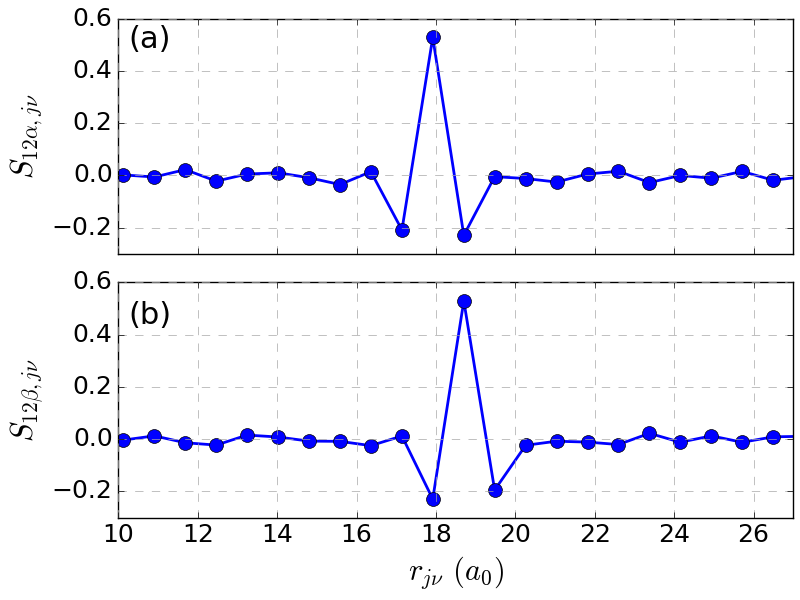}
 \caption{Spin-spin correlation functions plotted for the results obtained at $f=5\text{ Ry}/a_0$ and $N=50$ for both $\alpha$ (a) and $\beta$ (b) sites. As in Fig. \ref{fig:alc_ccc} the distinction between $\alpha$ and $\beta$ is arbitral since $a/2\approx b$.} 
 \label{fig:alc_sss}
\end{figure}
Similarly as for density-density, we analyze the spin-spin correlation functions to inspect the magnetic order. Likewise, we do not find any clear indication of long-range spin order, since $S_{i\mu,j\nu}$ functions  also decay on the same distances (Fig. \ref{fig:sss} and \ref{fig:alc_sss}). 
\section{Summary and Conclusions}
\label{sec:Summary}
In this work, we have analyzed the case of uniform  compression of the molecular-hydrogen linear chain   by means of the \emph{Variational Monte-Carlo} method combined with the \emph{ab-initio} approach which determines the renormalized single-particle wave function in the correlated state. Thus we complement the \emph{benchmark} model of the atomic linear chain with the analysis of its stabilization under influence of the external force, starting from the linear arrangement of the molecules (MLC).
We investigated the possibility of \emph{dissociation} of the molecular chain into the atomic linear chain within available accuracy. The finite size scaling analysis provided stabilization of the atomic phase. However, we are far from claiming that the distortion is completely suppressed at finite force according to the numerical precision, applied wave-function ansatz, and the simplified form of the Hamiltonian. Despite this uncertainty we emphasize that
 particular system configuration and its electronic state is tuned only  by means of a single controllable external parameter -- the  force $f$. In that sense, by considering one-dimensional enthalpy $h$,  we provided thermodynamic  solution for the system at $T=0K$. As it is also sometimes postulated for the solid hydrogen phases to become metallic before occurence of the expected atomization, it is in the \emph{molecular state} by means of the band gap closure ~\cite{Garcia}. Therefore, we inspected charge energy gap of the \emph{molecular chain} in the vicinity of the arrangement referring to the atomic state to find out if it expose metallic properties. At attainable precision we observed vanishing $\Delta$ indicating the presence of a metallic state which coincides with the atomization of chain. The qualitative analysis of correlation functions allowed to deduce lack of non-trivial charge order and spin order.

\begin{figure}
 \includegraphics[width=\linewidth]{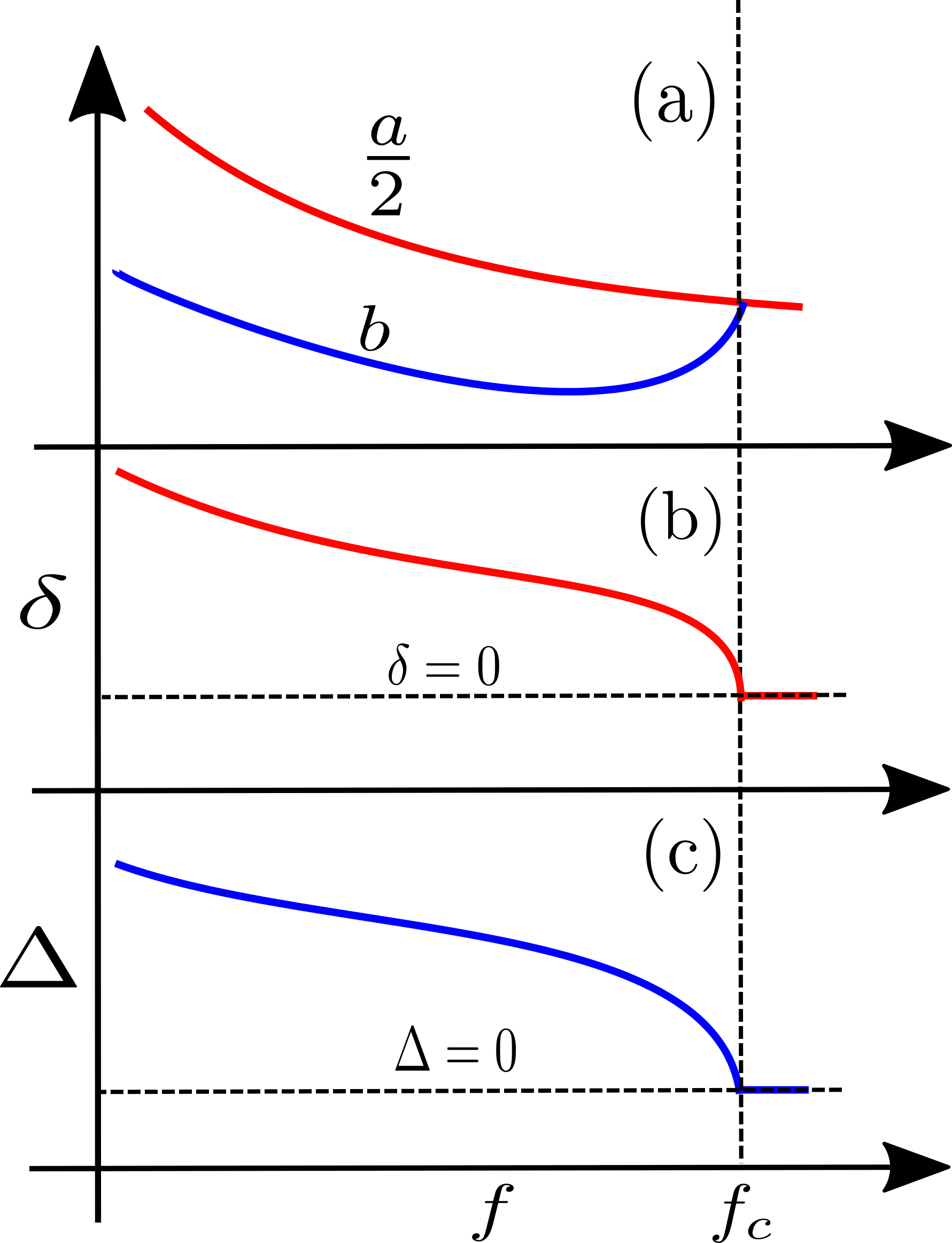}
 \caption{Schematic representation of the essential results.} 
 \label{fig:essence}
\end{figure}

\begin{figure}
 \includegraphics[width=\linewidth]{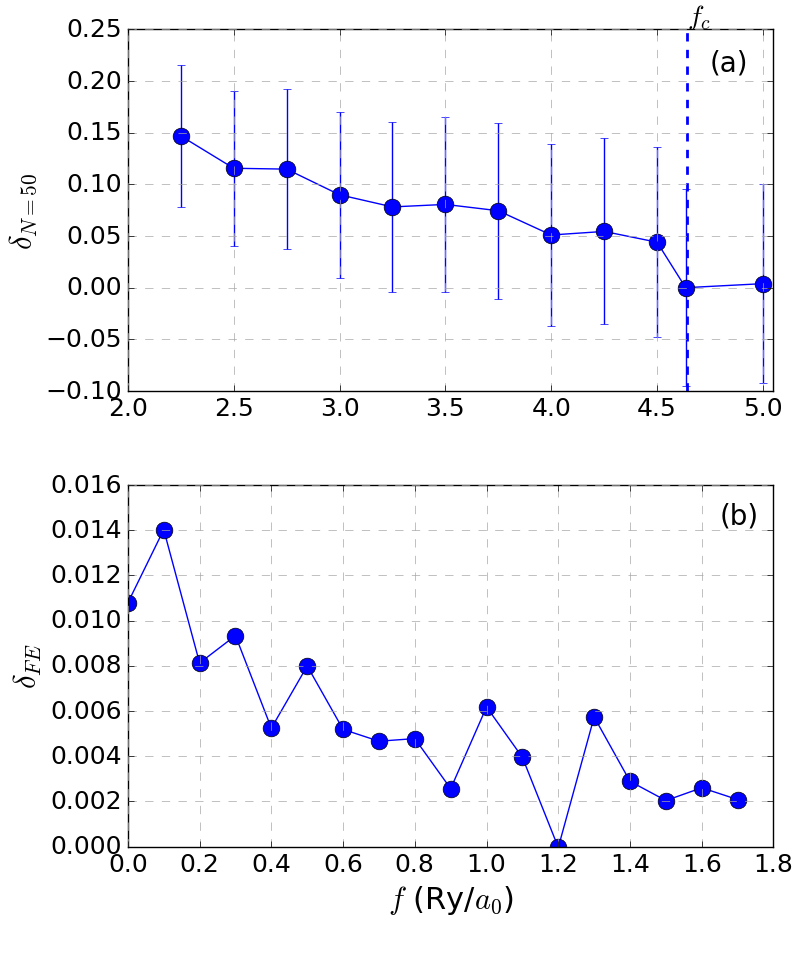}
 \caption{Distortion evolution parameter obtained by means of the VMC and with $N=50$ (a) and for the non-interacting electrons (b).} 
 \label{fig:distortion}
\end{figure}

The role of an external force f is crucial. Previously, we analyzed the ladder-type stacking of $H_2$ molecules ~\cite{Kadzielawa} and have shown that such a lateral arrangement is energetically stable even for $f=0$. This is not the case here and physically the role of the force $f$ may be played by a substrate with the chain placed on its surface. In the situation, in which the substrate lattice parameter is commensurate with the intermolecular distance, we can regard the force $f$ as a uniform compressing action on the chain. In that situation, a variable force could appear by changing either the substrate parameter or studying system on different substrates. Our analysis allows also for the Peierls distortion evolution via parameter 
\begin{align}
\label{eq:peierls_param}
\delta\equiv\frac{a-2b}{a}.
\end{align}
Namely, one sees that the correlations enhance the distortion in the molecular state, but it practically ceases to exist at the atomization (metallic) limit. Our results thus provide a systematic study of the distortion stability. Only a very small, residual value of $\delta \neq 0$ remains in the metallic phase. The two latter results are represented in Figs. ~\ref{fig:essence} and ~\ref{fig:distortion}, respectively. Explicitly, in Fig. ~\ref{fig:essence} we present schematically together the $a/2$ and $b$ distances, the Peierls distortion $\delta$, and the charge gap, defined by Eq. (~\ref{eq:energy_gap}). Those quantities characterize the atomization (a) at $f=f_c$, the associated with it disappearance of appreciable Peierls distortion, mainly caused by the chemical bonding in the molecular state (b), and disapperance of the charge gap at that point (c). All these characteristics, in conjuction with behavior of the density and spin correlations (cf. Figs.~\ref{fig:ccc}-~\ref{fig:alc_sss}), show that the atomization takes place to a standard-type of metallic state. This conclusion has been also reached in our analysis of metallization of $H_2$ ladders ~\cite{Kadzielawa} that the atomic phase is close to a moderately, if not weakly, correlated and thus standard, metallic. However, this means that the role of the electron-lattice interaction in the atomized state may become very important, as stated in a number of recent papers (see e.g. ~\cite{Borinaga2}), but this subject will not be discussed any further here.
Finally, one may ask a basic question, how the ordinary Peierls distortion picture fits into the above  picture, since the results depicted in Fig. ~\ref{fig:essence}b do not show any $\delta\neq0$ for $f\geq f_c$? To addres this question, we have plotted concrete data for $\delta$ in the correlated (cf. Fig. ~\ref{fig:distortion}a) and non-interacting (cf. Fig. ~\ref{fig:distortion}b) cases, respectively. One sees that spontaneous Peierls distortion $\delta_{FE}$ is about one order of magnitude smaller than that ($\delta$) in the correlated state. The $\delta_{FE}$ is practically on the border of numerical accuracy with increasing value of $f$.
\\
\indent Whereas we believe that we included remarkable part of electronic correlations in our model, we address the necessity to cover remaining matrix elements in the Hamiltonian e.g. correlated hoppings, exchange amplitudes or even three- of four-centres integrals to answer if provided conclusions are undoubtedly valid. Moreover, further inclusion of lattice dynamics and application of electron--phonon coupling may provide valuable outcome both in view of computational (\emph{benchmark}) aspects and physical mechanisms in the such phenomenon as conjectured room temperature  superconductivity in the metallic hydrogen ~\cite{Ashcroft,Szczesniak}.

\section*{Acknowledgments}
We thank Adam Rycerz and Michał Zegrodnik for many stimulating and clarifying discussions. 
The work was financially supported by the National Science Centre (NCN), through Grant {No.~DEC-2012/04/A/ST3/00342}.
\appendix
\section{Microscopic parameters}
In Tables~\ref{tab:hoppings} and ~\ref{tab:interactions} we provide all principal microscopic parameters (i.e., $t_{ij}^{\mu\nu},\epsilon,K_{ij}^{\mu\nu}$ and $U$), numbered as in Fig.~\ref{fig:model} for the range of forces considered in Fig. ~\ref{fig:aR50} for $N=50$. Note that $f=4.64 (\text{Ry}/a_0)$ and $f=5 (\text{Ry}/a_0)$ refer to a nearly \emph{atomic phase}. The hopping amplitudes between $\alpha$ and $\beta$ sites are negative, whereas those between sites from the same sublattice are positive. In  the atomic phase ($f \geq 4.64 (\text{Ry}/a_0) $), $t_1\approx t_5$, and $t_3\approx t_4$, as follows from the symmetry of the atomic system. Similar relations hold for the interaction parameters, i.e., $K_1\approx K_5$ and $K_3\approx K_4$. Note that although $U$ value is the highest, the inter-site interactions are up to $\sim 2/3$ of $U$, which means that we can easily have the situation with $2K\approx U$, which drives additionally the molecular system towards metallization, in addition to the single-particle energy.

\begin{table*}[t!]
\caption{The optimized hopping integrals and atomic energy obtained for $N=50$.\\ \\}
\label{tab:hoppings}
 \begin{tabular}{|c|c|c|c|c|c|c|c|}  $f$ (Ry/$a_0$) & $t_0$ (Ry) & $t_1$ (Ry) & $t_2$ (Ry) & $t_3$ (Ry) & $t_4$ (Ry) & $t_5$ (Ry)  & $\epsilon$ (Ry) \\\hline \hline
2.25  & 0.0820379  & -0.167832  & 0.67744  & -2.02135  & -2.45895  & -0.27246  &  -5.59409 \\\hline     
2.5  & 0.105143  & -0.215954  & 0.807776  & -2.42219  & -2.79632  & -0.316906  &  -5.54753 \\\hline     2.75  & 0.114176  & -0.231839  & 0.86109  & -2.57243  & -2.957  & -0.339721  &  -5.50698 \\\hline
3  & 0.125698  & -0.262781  & 0.928341  & -2.80194  & -3.1156  & -0.355215  &  -5.45454 \\\hline
3.25  & 0.13296  & -0.281589  & 0.971221  & -2.94012  & -3.21935  & -0.366422  &  -5.41488 \\\hline
3.5  & 0.143846  & -0.300785  & 1.0376  & -3.12185  & -3.42089  & -0.395242  &  -5.33852 \\\hline
3.75  & 0.147344  & -0.310661  & 1.05893  & -3.19222  & -3.47167  & -0.400123  &  -5.31453 \\\hline
4  & 0.157736  & -0.344547  & 1.12335  & -3.4169  & -3.61479  & -0.410421  &  -5.23649 \\\hline
4.25  & 0.167423  & -0.362063  & 1.18503  & -3.58491  & -3.80312  & -0.436923  &  -5.14978 \\\hline
4.5  & 0.175007  & -0.384558  & 1.23421  & -3.741  & -3.92157  & -0.448203  &  -5.08056 \\\hline
4.635  & 0.184093  & -0.43643  & 1.2946  & -4.00422  & -4.00438  & -0.437029  &  -4.99011 \\\hline
5  & 0.188949  & -0.444742  & 1.32733  & -4.08934  & -4.1061  & -0.451405  &  -4.9376 
\end{tabular}
\end{table*}

\begin{table*}[t!]
\caption{Electron-electron interaction integrals considered in the optimized state $N=50$.\\ \\}
\label{tab:interactions}
 \begin{tabular}{|c|c|c|c|c|c|c|c|}  $f$ (Ry/$a_0$) & $K_0$ (Ry) & $K_1$ (Ry) & $K_2$ (Ry) & $K_3$ (Ry) & $K_4$ (Ry) & $K_5$ (Ry)  &$U$ (Ry) \\\hline \hline
2.25  & 0.478449  & 0.627356  & 0.934136  & 1.56622  & 1.63987  & 0.641265  &  2.61841 \\\hline
2.5  & 0.514304  & 0.675894  & 1.00048  & 1.67231  & 1.72609  & 0.686352  &  2.72896 \\\hline
2.75  & 0.527862  & 0.693729  & 1.02576  & 1.71076  & 1.76321  & 0.704012  &  2.77628 \\\hline
3  & 0.544872  & 0.717087  & 1.05719  & 1.76318  & 1.80322  & 0.725021  &  2.8324 \\\hline
3.25  & 0.555068  & 0.73093  & 1.07595  & 1.79338  & 1.82756  & 0.737749  &  2.86468 \\\hline
3.5  & 0.570171  & 0.750606  & 1.10399  & 1.83494  & 1.86954  & 0.757562  &  2.91719 \\\hline
3.75  & 0.575004  & 0.757181  & 1.11288  & 1.84936  & 1.88109  & 0.763578  &  2.93283 \\\hline
4  & 0.589371  & 0.776929  & 1.1393  & 1.89326  & 1.91453  & 0.781249  &  2.97984 \\\hline
4.25  & 0.60242  & 0.793848  & 1.16348  & 1.92871  & 1.95109  & 0.798424  &  3.02543 \\\hline
4.5  & 0.612516  & 0.80744  & 1.18195  & 1.95763  & 1.97542  & 0.811095  &  3.05687 \\\hline
4.635  & 0.624954  & 0.825457  & 1.20478  & 1.99996  & 1.99997  & 0.825457  &  3.09791 \\\hline
5  & 0.631417  & 0.833722  & 1.21668  & 2.01667  & 2.01819  & 0.834038  &  3.11952 
\end{tabular}
\end{table*}

\section{Jastrow Variational parameters}

For the sake of completness we also present Jastrow variational parameters (cf. Tab.~\ref{tab:jastrows}) numbered in the similar manner as microscopic parameters. As expected their amplitudes corresponds directly to the magnitude of the interaction parameters. 

\begin{table*}[t!]
\caption{The Jastrow wave-function variational parameters for $N=50$ numbered in the same manner as the  microscopic parameters, cf. Tab. ~\ref{tab:hoppings}.\\ \\}
\label{tab:jastrows}
 \begin{tabular}{|c|c|c|c|c|c|c|c|}  $f$ (Ry/$a_0$)& $\lambda_0$ & $\lambda_1$ & $\lambda_2$ & $\lambda_3$ & $\lambda_4$ & $\lambda_5$  &$\lambda_U$ \\\hline \hline
2.25  & 0.0494324  & 0.0857595  & 0.151032  & 0.233874  & 0.226763  & 0.0906117  &  0.543334 \\\hline
2.5  & 0.0510948  & 0.0920228  & 0.161314  & 0.246269  & 0.241465  & 0.0961716  &  0.567389 \\\hline
2.75  & 0.0437915  & 0.0770094  & 0.131836  & 0.205208  & 0.200832  & 0.0786332  &  0.479059 \\\hline
3  & 0.0488745  & 0.0915902  & 0.16175  & 0.244208  & 0.242687  & 0.0957701  &  0.562443 \\\hline
3.25  & 0.0447105  & 0.082466  & 0.144809  & 0.221195  & 0.219908  & 0.0852848  &  0.513539 \\\hline
3.5  & 0.0419956  & 0.0788773  & 0.141055  & 0.217371  & 0.216265  & 0.0813463  &  0.504218 \\\hline
3.75  & 0.0442797  & 0.0830695  & 0.143916  & 0.218546  & 0.215457  & 0.0844753  &  0.502372 \\\hline
4  & 0.0432848  & 0.0814832  & 0.140217  & 0.21298  & 0.212156  & 0.0831675  &  0.491849 \\\hline
4.25  & 0.0430579  & 0.0815563  & 0.138759  & 0.210457  & 0.208788  & 0.080752  &  0.484049 \\\hline
4.5  & 0.0415801  & 0.0789011  & 0.135518  & 0.204778  & 0.205013  & 0.0795609  &  0.472317 \\\hline
4.635 & 0.0396584  & 0.0752711  & 0.128925  & 0.195331  & 0.195577  & 0.0755712  &  0.450903 \\\hline
5  & 0.0423223  & 0.080441  & 0.13515  & 0.202073  & 0.201048  & 0.0800144  &  0.461985
\end{tabular}
\end{table*}

\clearpage
\bibliography{bibliography}

\end{document}